\documentclass[aps,prb,preprint,superscriptaddress,showpacs]{revtex4}

\usepackage{graphicx}

\begin{document}

\def\Sr327{Sr$_3$Ru$_2$O$_7$}


\title{Multiple regions of quantum criticality in YbAgGe}




\author{G.~M. Schmiedeshoff}

\affiliation{Department of Physics, Occidental College, Los Angeles, California, 90041.}

\author{E.~D. Mun}

\affiliation{Ames Laboratory and Department of Physics and Astronomy, Iowa State University, Ames, Iowa, 50011.}

\author{A.~W. Lounsbury}

\affiliation{Department of Physics, Occidental College, Los Angeles, California, 90041.}

\author{S.~J. Tracy}

\affiliation{Department of Physics, Occidental College, Los Angeles, California, 90041.}

\author{E.~C. Palm}

\affiliation{National High Magnetic Field Laboratory, Florida State University, Tallahassee, Florida 32310.}

\author{S.~T. Hannahs}

\affiliation{National High Magnetic Field Laboratory, Florida State University, Tallahassee, Florida 32310.}

\author{J.~-H. Park}

\affiliation{National High Magnetic Field Laboratory, Florida State University, Tallahassee, Florida 32310.}

\author{T.~P. Murphy}

\affiliation{National High Magnetic Field Laboratory, Florida State University, Tallahassee, Florida 32310.}

\author{S.~L. Bud'ko}

\affiliation{Ames Laboratory and Department of Physics and Astronomy, Iowa State University, Ames, Iowa, 50011.}

\author{P.~C. Canfield}

\affiliation{Ames Laboratory and Department of Physics and Astronomy, Iowa State University, Ames, Iowa, 50011.}


\date{\today}

\begin{abstract}  

Dilation and thermopower measurements on YbAgGe, a heavy-fermion antiferromagnet, clarify and refine the magnetic field-temperature ($H$-$T$) phase diagram and reveal a field-induced phase with $T$-linear resistivity.  On the low-$H$ side of this phase we find evidence for a first-order transition and suggest that YbAgGe at 4.5 T may be close to a quantum critical end point.  On the high-$H$ side our results are consistent with a second-order transition suppressed to a quantum critical point near 7.2 T. We discuss these results in light of global phase diagrams proposed for Kondo lattice systems.

\end{abstract}



\maketitle

When a classical second-order phase transition is suppressed to zero temperature $T$ by a tuning parameter (such as pressure, doping, or magnetic field $H$), a quantum critical point (QCP) can occur \cite{Sachdev1999}; suppression of a first-order phase transition can lead to a quantum critical end point (QCEP) \cite{Millis2002}.  In the vicinity of a QCP or QCEP quantum fluctuations associated with the zero-point energies of the adjacent $T=0$ phases can persist to remarkably high temperatures.  These fluctuations can dramatically affect the interactions between particles, leading to unusual thermodynamic and transport properties \cite{Sachdev1999,Stewart2001,Millis2002,Lohneysen2007} and to novel states of matter \cite{Coleman2005,Gegenwart2008a,Schofield2010}. 

YbAgGe, a stoichiometric heavy-fermion (HF) antiferromagnet, crystallizes in a hexagonal ZrNiAl-type structure \cite{Poettgen1997}.  The zero-field electronic specific heat coefficient falls in the range 0.15-1.0 J/mol K$^2$ and the Kondo temperature is 20-25 K \cite{Budko2004a}. The Yb ions form a quasi-Kagom\'{e} lattice in which magnetic coupling, geometric frustration and Kondo interactions compete in a manner which allows the suppression of low temperature AF order by modest applied magnetic fields, fields that tune the quantum critical behavior \cite{Si2010a,Kim2010} and lead to a complex magnetic phase diagram \cite{Gignoux2001}.

\begin{figure}[htb]
\center{\includegraphics[height=4.0 in]{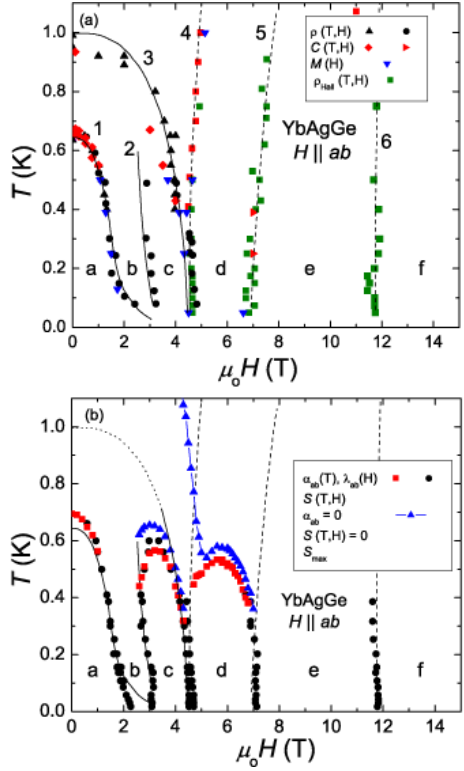}
\caption{(color online) The phase diagram of YbAgGe below 1 K with $H$ applied perpendicular to the $c$-axis. Panel (a) summarizes measurements from the literature, the variables carry their usual meanings. Panel (b) summarizes dilation and thermoelectric power data from this work with the same guides to the eye as panel (a) (see text).  \label{fig01}}}
\end{figure} 

The phase diagram of YbAgGe, summarizing earlier measurements \cite{Budko2008a} with $H$ perpendicular to the $c$-axis, is shown in Fig. \ref{fig01}a.  Solid (dashed) lines are guides to the eye for thermodynamic phase boundaries (`Hall lines' denoting features in the field-dependent Hall resistivity \cite{Budko2005a,Budko2005b}) labeled by numerals 1-3 (4-6), lower case letters $a$-$f$ label phases or regions of the phase diagram. Commensurate AF order is observed in the $a$-phase \cite{Fak2005a} where a sharp first-order phase transition manifests along phase line 1 \cite{Budko2004a}, and incommensurate AF order has been reported in the $b$-phase  \cite{Fak2006a}. Relatively broad features manifest along the higher temperature part of phase line 3 \cite{Budko2004a}, but below about 0.5 K the features sharpen and neutron scattering measurements reveal a return to commensurate AF order \cite{McMorrow2008}.  Based on the suppression of phase line 3 near 4.5 T and its near coincidence with Hall line 4 below 0.3 K, a field-induced QCP was proposed \cite{Budko2004a,Budko2005a,Budko2005b}.  Intriguingly, several other features of this phase diagram remained mysterious.

\begin{figure}[htb]
\includegraphics[height=4.0 in]{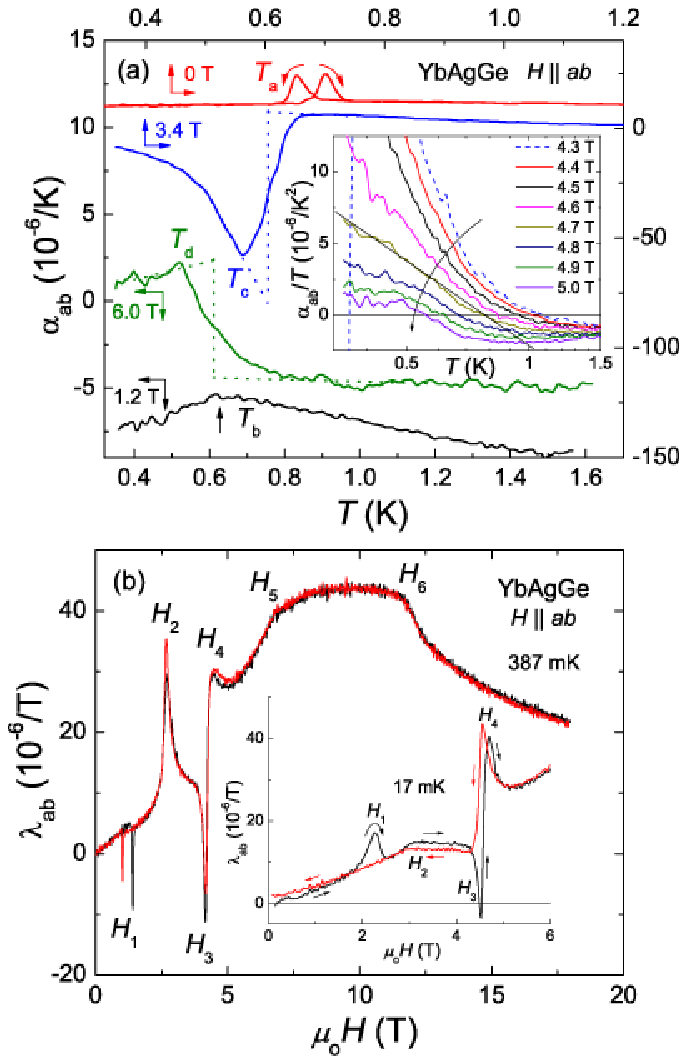}
\caption{(color online) Representative longitudinal dilation data with $H$ applied parallel to the $ab$-axis.  Panel (a) shows the low temperature thermal expansion in the fixed fields listed next to arrows denoting the appropriate axes for each data set (the data at 0 T/1.2 T have been shifted by $+10/-9.0\times{10}^{-6}$/T for clarity).  The $T_i$ label features associated with the phase boundaries or crossovers of Fig. 1 (see text).  A semi-log plot of $\alpha_{ab}/T$ vs. $T$ in the vicinity of 4.5 T is shown in the inset where the direction of increasing $H$ is shown by a curved arrow and the solid line is a guide to the eye for the low temperature 4.7 T data.  Panel (b) shows the magnetostriction at 387 mK.  The $H_j$ label features in the data where $j$ denotes the guides to the eye of Fig. 1a (see text).  The inset shows an expanded view of the transitions below 6 T at 17 mK. \label{fig02}} 
\end{figure}

The low-$T$ electrical resistivity is large \cite{Budko2004a} and varies like $T^n$ with $n\simeq{1}$ in region $d$, smoothly increasing from $\simeq{1}$ to $\simeq{2}$ in region $e$, and $\simeq{2}$ in region $f$ \cite{Niklowitz2006}. Such non-Fermi liquid (nFL) behavior is expected near a QCP \cite{Stewart2001,Lohneysen2007}, but the broad field range with $n\simeq{1}$ and the recovery of Fermi liquid (FL) behavior in fields so far above that of the QCP are surprising.  Further, a logarithmic divergence of the specific heat appears most clearly for $H\sim$ 7 T \cite{Budko2004a}, near Hall line 5, raising the possibility of (at least) one other phase being suppressed in fields well above that of the proposed QCP near 4.5 T.  In this paper we describe longitudinal dilation and transverse thermoelectric power (TEP) measurements on YbAgGe that shed light on some of these mysteries and paint a fuller picture of quantum criticality in Yb-based HFs.

\begin{figure}[htb]
\includegraphics[height=4.0 in]{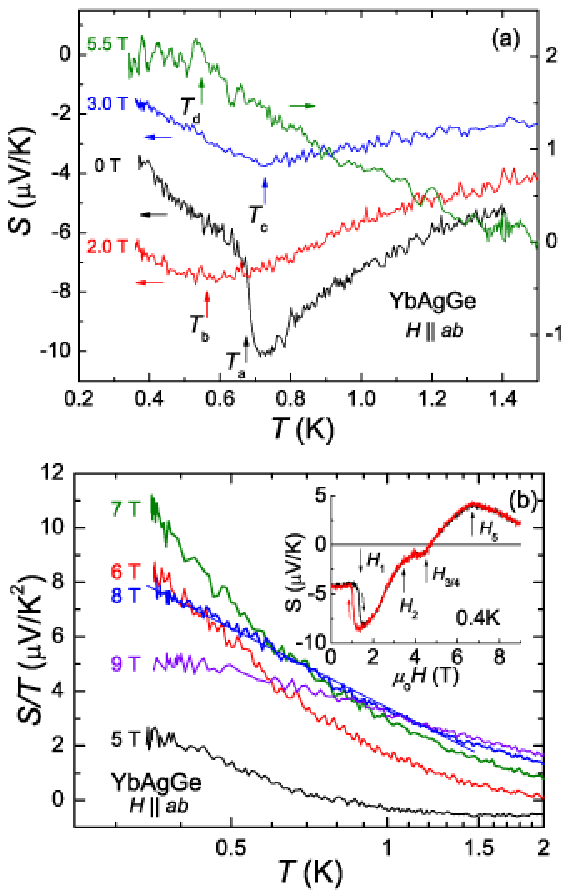}
\caption{(color online) Representative transverse TEP data with $H$ applied parallel to the $ab$-axis and the temperature gradient in the basal plane, perpendicular to $H$.  Panel (a) shows the temperature dependence of $S$ across phase boundaries or crossover regions at the fields shown.  The $T_i$ label features associated with the phases (or regions) of Fig. 1. Panel (b) shows $S/T$ vs. $T$ in applied fields on a semi-log plot.  The solid line through the 8 T data is a guide to the eye.  The field dependence of $S$ at 0.4 K is shown in the inset where the $H_i$ label features in the data and $i$ denotes the guides to the eye of Fig. 1a. \label{fig03}}  
\end{figure}

Single crystals of YbAgGe were grown from an AgGe-rich ternary solution \cite{Budko2004a}.  Longitudinal dilation and transverse TEP measurements were made with techniques described elsewhere \cite{Schmiedeshoff2006b,Mun2010a}. $H$ was applied normal to the $c$-axis.  The coefficients of linear thermal expansion and linear magnetostriction along the $ab$-axis (approximately parallel to [210]) are $\alpha_{ab} = \partial(\ln{L_{ab}})/\partial{T}$ and $\lambda_{ab} = \partial(\ln{L_{ab}})/\partial{H}$ respectively where $L_{ab}$ is the thickness of the sample, 1-2 mm.  The Seebeck coefficient is $S = -\Delta{V}/\Delta{T}$, where $\Delta{V}$ is the potential difference across the sample when a temperature gradient is applied. Typical data are shown in Figs. \ref{fig02} and \ref{fig03} where nFL behavior is also illustrated.

In our dilation data, first-order phase transitions are identified by their peak-like shape and the presence of thermal or magnetic hysteresis ({\it e.g.} features labeled $T_a$ and $H_{1-4}$ in Fig. 2); second-order transitions do not show hysteresis within experimental uncertainty and usually show a step-like shape as illustrated by the dashed lines through the 3.4 and 6.0 T data of Fig. 2a.  Pronounced features in the 3.4 and 6.0 T data of Fig. 2a, labeled $T_{c,d}$ correlate well with extrema in $\lambda_{ab}$ and are used to construct the phase diagram.  The first-order transitions observed in the magnetostriction exhibit varying levels of magnetic hysteresis.  This hysteresis becomes quite pronounced as the temperature falls below about 0.2 K, as illustrated in the inset of Fig. 2b where the features associated with $H_1$ and $H_3$ change dramatically in the $dH/dt < 0$ data.  For all of our $T \leq$ 0.2 K magnetostriction data the sample was zero-field-cooled from temperatures above 0.3 K. Fig. 1b was constructed from $dH/dt > 0$ data. 

The phase diagram assembled from our dilation and TEP measurements is shown in Fig. 1b where the solid and dashed lines are those of Fig. 1a, except for the dotted portion of phase line 3 where we observe no thermodynamic features.  In comparison with the phase diagram of Fig. 1a we find: that phase line 2 joins phase line 3 to surround the $c$-phase; that Hall lines 4-6 are all associated with thermodynamic phase transitions as $T \rightarrow 0$; and that phase transitions anchoring Hall lines 4 and 5 define the low temperature boundaries of a field-induced $d$-phase (the high temperature limits of the $n\simeq{1}$ region of the resistivity agree reasonably well with the top of the $d$-phase dome \cite{Niklowitz2006}). The phase transitions are second-order across the tops of the $c$- and $d$-phase domes and first-order along phase line 3 (Hall line 4) below 0.3 K (0.2 K).  The overlap of the phase line data determined from $\alpha_{ab}$, $\lambda_{ab}$, and $S(T,H)$ is extensive along the $c$-phase boundary, there is some overlap on the high-field side of the $d$-phase but no overlap is observed on the low-field side where a small gap appears between the $\alpha_{ab}$ and $\lambda_{ab}$ data, a gap Hall line 4 passes through.  Broad extrema in $\alpha_{ab}(T)$ and $S(T)$, labeled with $T_b$ in Figs. \ref{fig02}a and \ref{fig03}a respectively, may represent a cross-over region between higher-$T$ fluctuations and the lower-$T$ incommensurate AF order of the $b$-phase, leading to the uncertain nature of the phase diagram in this region.

Several signatures of quantum criticality appear in our data.  A change in the sign of the thermal expansion at a QCP has been predicted \cite{Garst2005}, the phase diagram coordinates where $\alpha_{ab}$ passes through zero are plotted as solid triangles in Fig. \ref{fig01}b.  These `lines of zeros' pass along the tops and high-$H$ sides of the $c$- and $d$-phases and presumably extend to $T = 0$ near 4.5 and 7.2 T.  Hall line 4 (5) correlates with the line of zeros associated with the $c$-phase ($d$-phase) as $T\rightarrow{0}$.  A sign change in $S(T)$ correlates with the field-induced QCP in YbRh$_2$Si$_2$ where it is attributed to an abrupt change in the Fermi surface \cite{Hartmann2010a}, though such a sign change does not appear to be a universal QCP signature \cite{Kim2010}.  $S/T$ as $T \rightarrow 0$ is predicted to reach its maximum value as the QCP is approached and its symmetry with respect to the QCP can help distinguish between theoretical models \cite{Kim2010}.  Sign changes in $S(T,H)$ for YbAgGe are plotted as solid diamonds in Fig. 1b and correlate with Hall line 4. Maxima in $S(H)$ are plotted as solid stars in Fig. 1b and correlate with Hall line 5 where the largest values of $S/T$ as $T \rightarrow 0$ also appear.

\begin{figure}[htb]
\includegraphics[width=2.9 in]{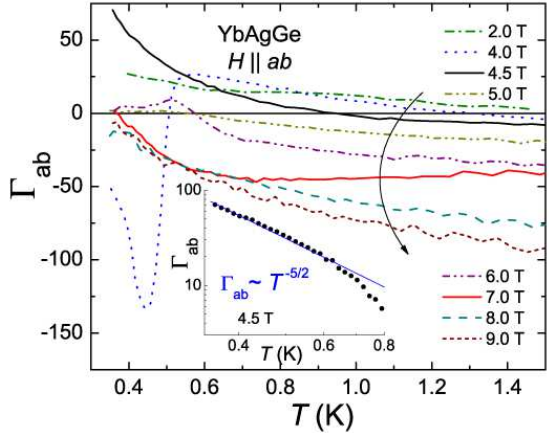}
\caption{(color online) The temperature dependence of the Gr\"{u}neisen parameter $\Gamma_{ab}$ of YbAgGe in applied magnetic fields. The curved arrow represents the direction of increasing field.  (The apparent, slight downturn of the 8.0 T data at the low-$T$ end is comparable to the noise in these data.) The inset shows the low temperature end of the 4.5 T data on a log-log plot.  The solid line is a guide to the eye representing a $T^{-5/2}$ temperature dependence. \label{fig04}}  
\end{figure}

The Gr\"{u}neisen parameter, characterizing the volume dependence of the energy scales in the system (which should be dominated by the quantum critical contribution as $T \rightarrow 0$), is defined as $\Gamma_{ab} = V_m\alpha_{ab}/\kappa{C_p}$ where $V_m$ is the molar volume \cite{Poettgen1997}, $\kappa$ is the compressibility \cite{Sengupta2010}, and $C_p$ is the specific heat \cite{Tokiwa2006}. $\Gamma_{ab}$ is predicted to diverge as a QCP is approached \cite{Zhu2003}, our results are shown in Fig. 4.  The largest values of $\Gamma_{ab}$, comparable to that of other HF compounds \cite{deVisser1990}, occur at 4.5 T.  At this field the low-$T$ upturn in $\Gamma_{ab}(T)$ is consistent with the onset of Gr\"{u}neisen divergence, perhaps with a power-law temperature dependence as suggested by the data in the inset of Fig. 4 (though the temperature range over which divergent behavior is observed is too limited, at this time, for a definitive quantitative analysis).  $\Gamma_{ab}(T)$ at 7, 8, and 9 T also show low-$T$ upturns suggesting that Gr\"{u}neisen divergence may develop at lower temperatures; we cannot, as yet, explain the large magnitudes nor the shapes of $\Gamma_{ab}(T)$ above 0.5 K in this field range.  It will be necessary to extend these measurements to lower temperatures, higher fields, and at a higher density of field values.

We take the presence of nFL behavior (both earlier \cite{Budko2008a} and current: the  logarithmic temperature dependences of $\alpha_{ab}/T$ and $S/T$ suggested by the data in the inset of Fig. 2a and the main panel of Fig. 3b respectively), the features in the Hall resistivity \cite{Budko2005a,Budko2005b} (Hall lines 4 and 5), the zeros in $\alpha_{ab}(T)$, and the features and zeros in $S(T,H)$ as strong evidence for (at least) two regions of quantum criticality in YbAgGe near $H_{c1} = 4.5$ T (also supported by the onset of Gr\"{u}neisen divergence) and $H_{c2} = 7.2$ T.  

If a QCP at $H_{c1}$ is due to the suppression of the AF transition characterized by the full phase line 3, as previously suggested \cite{Budko2004a}, then one would expect a continuous transition as $T \rightarrow 0$ \cite{Sachdev1999}, perhaps similar to that observed in YbRh$_2$Si$_2$ \cite{Gegenwart2008a}.  However, the transitions along phase line 3 are first-order (see Fig. 2b) as $T \rightarrow 0$. A step-like feature developing in the low-$T$ magnetization near $H_{c1}$ \cite{Tokiwa2006} and the thermodynamic structure of the $c$- and $d$-phases (two second-order phase lines approaching/joining a first-order phase line extending to $T=0$) are similar to that expected for the spin-flop class of metamagnetism \cite{Stryjewski1977,Aharony1983}.  Metamagnets can exhibit a wide variety of multicritical behavior, including intermediate anisotropy scenarios leading to critical end points \cite{Vilfan1986}.  We suggest that YbAgGe may be close to a QCEP near $H_{c1}$, possibly similar to that observed in \Sr327 \cite{Grigera2001a} but with spin-flop metamagnetism driving the quantum criticality.

The phase transitions at the base of Hall line 5 are clearly continuous as shown in the main panel of Fig. 2b while $n\simeq{1}$ and $\alpha_{ab}(T)$ passes through zero here.  In this region of the phase diagram both the specific heat  and $S/T$ are logarithmically divergent.  We propose that a QCP occurs near $H_{c2} = 7.2$ T, driving the pronounced nFL behavior in this region of the phase diagram.

Theoretical efforts, characterizing several quantum critical materials on a global phase diagram incorporating Kondo coupling and degree of magnetic frustration, suggest that YbAgGe may evolve from AF order ($d$-phase) through a spin-liquid phase ($e$-phase) before a FL ($f$-phase) is recovered \cite{Si2010a,Custers2010a}.  Recent work on other Yb-based HF compounds suggests \cite{Custers2010a,Coleman2010} that a quantum critical {\it phase}, bounded by QCPs at $H_{c2}$ and a hypothetical spin-liquid/heavy FL QCP \cite{Si2010a} near 12 T, may underlie the spin-liquid.  The linear $T$-dependence and large magnitude of the resistivity suggest strange-metal behavior for the $d$-phase \cite{McGreevy2010}, and as a spin-flop phase it will carry a net magnetic moment \cite{Stryjewski1977}, though the small step in $M(H)$, about $0.1 \mu_B/$Yb \cite{Tokiwa2006}, suggests that an underlying AF symmetry may still be present. The theoretical identification of the $e$-phase as a spin-liquid is supported by the large magnetostriction \cite{Ruff2010} we observe. The proposed global phase diagram thus seems appropriate for YbAgGe.  Elastic and inelastic neutron scattering would be profitable though challenging microscopic probes of these high-$H$ phases.  

In conclusion, dilation and TEP measurements reveal high-$H$ phase boundaries in YbAgGe that delineate the region of $T$-linear resistivity.  On the low-$H$ side this phase appears to be close to a QCEP near 4.5 T, associated with a first-order, most likely metamagnetic, phase transition. On the high-$H$ side this phase appears to end in the continuous suppression of a second-order transition ending in a QCP near 7.2 T, explaining the pronounced nFL behavior nearby.  Even with the identification of this field-stabilized phase, there remains a clear nFL region over which the resistivity varies as $T^n$ with $n$ continuously changing from $\simeq{1}$ to $\simeq{2}$ as $H$ increases.  Theory suggests a quantum critical phase and/or spin-liquid in this region but more evidence, theoretical as well as experimental, is needed.  

This work was supported by the National Science Foundation under DMR-1006118.  Work at Ames Laboratory was supported by the U.S. Department of Energy, Basic Energy Sciences, under Contract No. DE-AC02-07CH11358.  Work at the National High Magnetic Field Laboratory was supported under the auspices of the National Science Foundation, the State of Florida, and the U.S. Department of Energy.

\end{document}